\documentclass{mem}
\usepackage{natbib}\usepackage{balance}
\usepackage{graphicx}
\usepackage{savesym}
\usepackage{amsmath}
\savesymbol{iint}
\usepackage{txfonts}
\usepackage[a4paper,breaklinks,dvipdfm]{hyperref}
\restoresymbol{TXF}{iint}
\idline{1}{1}
\begin{document}
\def\teff{$T\rm_{eff }$}
\def\kms{$\mathrm {km s}^{-1}$}
\newcommand{\del}{\partial}
\newcommand{\xhat}{\mathbf{\hat{x}}}
\newcommand{\yhat}{\mathbf{\hat{y}}}
\newcommand{\Xhat}{\mathbf{\hat{x}_0}}
\newcommand{\Yhat}{\mathbf{\hat{y}_0}}
\newcommand{\vel}{\mathbf{v}}
\newcommand{\velA}{\mathbf{v_A}}

\title{
Propagation of Low-Energy Cosmic Rays in Molecular Clouds: Calculations in Two Dimensions
}

   \subtitle{}

\author{
P. Rimmer\inst{1} 
\and E. Herbst\inst{2}
          }

  \offprints{P. Rimmer}

\institute{
Ohio State University, Department of Physics --
191 W Woodruff Ave,
Columbus, Ohio 43219, USA
\email{pbrimmer@mps.ohio-state.edu}
\and
University of Virginia, Department of Chemistry --
McCormick Road, P.O. Box 400319, Charlottesville, VA 22904-4319 
}

\authorrunning{Rimmer }

\titlerunning{Cosmic Ray Propagation in Molecular Clouds}

\abstract{ We calculate cosmic ray transport with a collisional Boltzmann Transport Equation, including E-M forces. We apply the Crank-Nicholson Method to solve this equation. At each time step, the spatial distribution of cosmic rays is applied to the ZEUS 2D magnetohydrodynamics model, which is then utilized to calculate the resulting E-M field. Finally, the field is applied to the initial equation. This sequence is repeated over many time steps until a steady state is reached. We consider results from $t = 0$ until steady state for an isotropic low energy cosmic ray flux, and also for an enhanced cosmic ray flux impinging on one side of a molecular cloud. This cosmic ray flux is used to determine an ionization rate of interstellar hydrogen by cosmic rays, $\zeta$. Astrochemical implications are briefly mentioned.
}
\maketitle{}

\section{Introduction}

Low energy ($< 1$ GeV) cosmic rays drive interstellar chemistry and may cause specific spectral features recently measured, such as the 6.7 keV emission line. Yet the origin and flux of low energy cosmic rays is currently unknown because the Sun's magnetic field deflects these particles, so that they cannot be directly observed. There is a great deal of uncertainty about the correct cosmic-ray flux-spectrum for low energy cosmic rays, ranging from a steep slope of $\sim E^{-3}$ (predicting a great many low energy cosmic rays, see \cite{Nath1994} all the way to a positive slope of $\sim E$ \citep[][predicting very few low energy cosmic rays]{Spitzer1968}. A robust model of cosmic ray transport in molecular clouds is necessary in order to understand this flux-spectrum as a function of position within a molecular cloud. Modelling low energy cosmic ray streaming will afford better understanding of interstellar chemistry and possible line emissions caused by these cosmic rays. We present such a model here.

\section{Equation and Numerical Method}

We have absolute coordinates $\Xhat$ and $\Yhat$ defining the position within the cloud. For each of these values, and for various momenta, $p$, pointed in directions determined by angle $\mu$, we define a distribution function for cosmic rays, $f$. In the scheme utilized by \cite{Skilling1975a}, we solve $f$ using a semi-collisional relativistic Boltzmann Equation in a two-fluid approximation, where the cosmic rays are treated as one fluid, and the interstellar medium as another fluid. The equation is semi-collisional in that collisions within the interstellar medium are treated, as are collisions between cosmic rays and the medium, but not cosmic rays colliding with other cosmic rays.

In our two dimensional scenario, we set up a local coordinate system ($\xhat$,$\yhat$) for each small area, such that $\xhat \cdot \mathbf{B} = B$ and $\yhat \cdot \mathbf{B} = 0$. $\mathbf{B}$ is separated from $\Xhat$ by an angle $\alpha$ such that the simple rotational transformation will map $\Xhat$,$\Yhat$ $\rightarrow$ $\xhat$,$\yhat$. We then apply the Fokker-Planck equation to solve for the distribution function. We use the Fokker Planck equation of a form similar to that of \citet{Cesarsky1978}.
\begin{equation}
\begin{split}
\frac{df}{dt} + \frac{c\mu p}{\gamma}\frac{\del f}{\del x} - \frac{1 - \mu^2}{2}\frac{cp}{\gamma}\frac{\del \ln B}{\del x}\frac{\del f}{\del \mu} \\
= p \frac{\del f}{\del p} \bigg[ \mu^2 (\xhat \cdot \vel)\frac{\del f}{\del x} + \frac{1-\mu^2}{2}(\yhat \cdot \vel)\frac{\del f}{\del y} \bigg] \\
+ (1-\mu^2)\mu\frac{\del f}{\del \mu}\bigg[(\xhat \cdot \vel)\frac{\del f}{\del x} - \frac{1}{2}(\yhat \cdot \vel)\frac{\del f}{\del y} \bigg] \\
+ \bigg(\frac{dp}{dt}\bigg)_{\rm coll} \frac{\del f}{\del p} + \frac{\del}{\del \mu}\bigg(D_{\mu} \frac{\del f}{\del \mu}\bigg) - \frac{p}{2}\bigg(\frac{\del \mathbf{u}}{\del p}\bigg)\cdot \nabla f 
\end{split}
\label{eq:fokker}
\end{equation}
Now, $df/dt = \del f/\del t + (\vel\cdot\nabla)f$, $c$ the speed of light, $\gamma = (1 - v^2/c^2)^{-1/2}$, $D_{\mu}$ is the pitch angle diffusion coefficient, and $\mathbf{u}$ is the velocity of a mean wave frame, or $\vel + \velA$. The l.h.s. terms in Equation (\ref{eq:fokker}) describe the change in motion of relativistic charged particles due to convective motion of the plasma itself, as well as cosmic ray streaming along the magnetic field. The first two terms on the r.h.s. describe the effects of the change in plasma densities. The third term is momentum change due to inelastic and elastic two-body collisions with the medium, and the fourth term describes the scattering of particles due to two-body collisions as well as magnetic field irregularities. The last term on the r.h.s., and the largest change we have made to \cite{Cesarsky1978}, is an addition from \cite{Skilling1975a}. This term accounts for spatial diffusion across $\mathbf{B}$.

This equation does not include source terms or acceleration mechanisms, although they are relatively straight-forward to include both in Equation (\ref{eq:fokker}) and its numerical solution, so long as some already-established acceleration mechanism is provided.

For collisions, we separate the collisional momentum change into elastic and inelastic terms, referred by the subscripts ``in'' and ``el'', respectively, and the approximation for the inelastic case is:
\begin{equation}
\bigg(\frac{dp}{dt}\bigg)_{\rm in} \approx \frac{n\sigma_{\rm in} p}{\gamma m} \Delta p;
\label{eq:in_approx}
\end{equation}
where $\sigma_i$ is the inelastic scattering cross-section from \cite{Cravens1975} and other sources, listed and reviewed very well in \cite{Padovani2009}. The other terms, $\Delta p$ is the momentum change from each collision, also reviewed in \cite{Padovani2009} and \cite{Rimmer2011}. $n$ is the density of the cloud, and $m$ is the mass of the cosmic ray particle, either the electron or proton mass. Elastic scattering is dealt with in a similar manner, except that the momentum is conserved over the two bodies involved in the collision, and the scattering cross-section is different. It is important to note that the elastic scattering also impacts $D_\mu$.

We solve this equation using the Crank-Nicolson method \citep{Crank1947}, evolving the system from $x_i,y_i,p_i,\mu_i,t_i \rightarrow x_i,y_i,p_i,\mu_i,t_{i+1}$, and all iterations thereof. We approximate the first and second derivatives to, for example:
\begin{equation}
\begin{split}
\frac{\del f}{\del t} = \frac{f(x_i,y_i,p_i,\mu_i,t_{i+1}) - f(x_i,y_i,p_i,\mu_i,t_i)}{\Delta t} \\
\frac{\del^2 f}{\del \mu^2} = \frac{1}{2(\Delta x)^2} \bigg( f(x_i,y_i,p_i,\mu_{i+1},t_{i+1}) \\ 
- 2f(x_i,y_i,p_i,\mu_i,t_{i+1}) + f(x_i,y_i,p_i,\mu_{i-1},t_{i+1}) \bigg) \\
+ \bigg( f(x_i,y_i,p_i,\mu_{i+1},t_i) - 2f(x_i,y_i,p_i,\mu_i,t_i) \\
+ f(x_i,y_i,p_i,\mu_{i-1},t_i) \bigg);
\end{split}
\label{eq:elements}
\end{equation}
where $\Delta t$ and $\Delta x$ are characteristic time and length scales. For a cloud one parsec in diameter, with consideration for the constants in Equation (\ref{eq:fokker}), the $\Delta t$ must be set to less than $\sim 1$ year without becoming too inaccurate. What is advantageous about the Crank-Nicolson method is its stability. The method will not lose stability pretty-much regardless of the choice of length and time scales, and so it is relatively straight-forward to determine the self-consistent accuracy of the calculations.

All the values for $f$ for $x,y,p,\mu$ at $t = 0$ are given, as are the values for $f$ at the boundaries, $x = 0,L; y = 0,L$ over all values of $p,\mu$. In this proceeding, we use a set flux-spectrum to determine the value for $f$ at the boundaries, and this value does not change with time. At $t = 0$, the value for $f$ is determined by the flux-spectrum at the boundary, and $f = 0$ at all other points in the cloud.

The calculation proceeds from the initial conditions of $x_i,y_i,p_i,\mu_i,t_0 = 0$, applying the values in Equations (\ref{eq:elements}) to Equation (\ref{eq:fokker}), arranging the elements as a series of matrices, inverting these matrices, and solving for the unknown values of $f$ at $x_i,y_i,p_i,\mu_i,t_1$. This is continued until steady-staet is reached.

At the same time, the electromagnetic field and local density are determined using the ZEUS magnetohydrodynamics code \citep{Stone1992}. The input to the ZEUS code is a charge distribution provided by the cosmic ray distribution functions for electrons and protons. The results of the ZEUS code are applied repeatedly for each time step for the transport equation.

\section{Results in terms of the Ionization Rate}

It is useful for astrochemists, and also conceptually advantageous, to represent the two-dimensional results for the cosmic ray distribution in terms of a cosmic ray ionization rate, $\zeta$ which is the rate at which hydrogen atoms are ionized by cosmic rays. This can be achieved mathematically by converting the distribution function to a position-dependent flux-density, $j(x,y,p) = p^2 f(x,y,p)$. To derive a position-dependent ionization rate from the flux-density, we use the form from \cite{Spitzer1968} with constants in front to account for ionization caused by the products of the first ionization, $\chi_2$, discussed in \cite{Dalgarno1999}. This is the ionization rate for protons:
\begin{equation}
\xi_{p}(x,y) = 4\pi \chi_2 \int_{E_{\rm min}}^{\infty} j(x,y,E) \sigma_{i,p} dE.
\label{eq:crden}
\end{equation}
In this equation, $\sigma_i$ is the ionizing cross-section and $E_{\rm min}$ is the minimum energy for cosmic ray ionization. We achieve $\zeta$ by averaging $\xi$ along a line passing through the two-dimensional cloud.

We performed the calculations for $f(x,y,p,\mu)$. For both calculations, the flux at the boundary is taken from \cite{Nath1994}. For the first case, the flux is isotropic and there is a low-energy cutoff for the initial flux-density of $1$ MeV (of course, the flux density inside the cloud can extend down to $E_{\rm min}$). In the second case, the initial flux extends down to $E_{\rm min}$ but impinges only on one side. The other side has the same initial flux-density, but with the $1$ MeV cutoff.

For the first case, the cosmic ray ionization rate extends from about $7.5 \times 10^{-17}$ s$^{-1}$ at the center to $2 \times 10^{-16}$ s$^{-1}$ at the edges. This difference is too small to accurately detect, given that chemical tracers are the best current way to determine the cosmic ray ionization rate, and are accurate only to within a factor of 2 or 3 \citep[see][for a review] {McCall2003,Indriolo2007,LePetit2006}. In the second case, however, the ionization rate spans two orders of magnitude, and should definitely be within detection capability, provided that sources can be found near the sites of cosmic ray production and with angular resolution capable of achieving length-scales of about 10-100 AU.

\begin{figure}[]
\resizebox{\hsize}{!}{\includegraphics[clip=true]{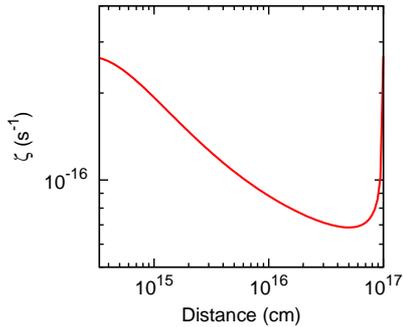}}
\caption{
\footnotesize
$\zeta$ as a function of depth into the cloud for an isotropic flux at the boundary from \cite{Nath1994} with a minimum energy of $1$ MeV.
}
\label{fig:zeta_isotropic}
\end{figure}

\begin{figure}[]
\resizebox{\hsize}{!}{\includegraphics[clip=true]{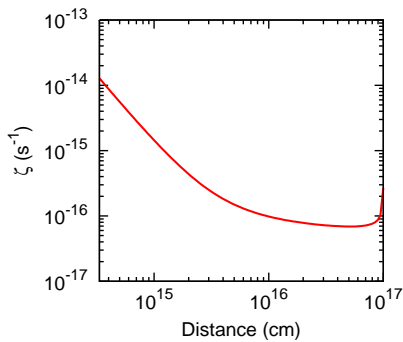}}
\caption{
\footnotesize
$\zeta$ as a function of depth into the cloud for an isotropic flux at the boundary from \cite{Nath1994} with a minimum energy of $1$ MeV on the right side, and $100$ eV on the left side.
}
\label{fig:zeta_anisotropic}
\end{figure}

\section{Discussion and Future Work}

To more thoroughly examine the ionization of cosmic rays, we need to treat electrons as well as protons. The cross-sections have already been included in the code, and the electron cosmic ray streaming will be calculated simultaneously with the proton cosmic rays as a logical next step. Eventually a third dimension and turbulence, as well as self-gravitation will be incorporated in the calculations.

There are many other questions such a model may answer beyond the cosmic ray ionization rate, such as what are the dominant magnetic effects on low energy cosmic rays. Candidates include magnetic mirroring\citep[discussed in][]{Cesarsky1978}, Alfv\'{e}n weaves\citep{Skilling1976}, and gravitational and turbulence-driven effects. Eventually, Fermi acceleration and shock-driven acceleration will be added to the model, so that the origin and range of these low energy cosmic rays can be theoretically explored.

The main problem that this code addresses now is the question of the cosmic ray ionization rate, and why it has the value that it does, connecting it with a flux-spectrum that depends on cloud geometry, composition, and physical properties like density and electromagnetic properties. At the end of his 2006 review, Alex Dalgarno stated that ``The interesting question may be not why are [cosmic ray ionization rates] so
different but why are they so similar \citep{Dalgarno2006}.'' The preliminary results of this study suggest that a combination of geometry and magnetic field effects may provide the answer to both questions.
 
\begin{acknowledgements}
We are exceptionally grateful to Alexandre Marcowith for organizing the conference and surrounding activities, and for his invitation to the beautiful city of Montpellier. P.B.R. also thanks the DeMartini Scholarship and Ohio State University for its financial support.
\end{acknowledgements}

\bibliographystyle{aa}

\end{document}